\documentclass[12pt]{iopart}   
\usepackage{iopams}
\usepackage{epsf}
\usepackage{euscript}
\usepackage{graphicx}

\renewcommand{\footnoterule}

\begin{document}
\title {Quantum chaos on discrete graphs}
\author{Uzy Smilansky } \address{Department of Physics of
Complex Systems, The Weizmann Institute of Science, Rehovot 76100,
Israel.\\
and\\
Isaac  Newton Institute for Mathematical Sciences, 20 Clarkson
Road, Cambridge CB3 0EH,\ UK.}
\date{\today}
\ead{uzy.smilansky@weizmann.ac.il}

\begin{abstract}
Adapting a method developed for the study of quantum chaos on {\it
quantum (metric)} graphs   \cite {KS}, spectral $\zeta$ functions
and trace formulae for {\it discrete} Laplacians on  graphs are
derived. This is achieved by expressing the spectral secular
equation in terms of the periodic orbits of the graph, and
obtaining  functions which belongs to the class of $\zeta$
functions proposed originally by Ihara \cite {Ihara}, and expanded
by subsequent authors  \cite {Stark,Sunada}. Finally, a model of
``classical dynamics" on the discrete graph is proposed. It is
analogous to the corresponding classical dynamics derived for
quantum graphs \cite {KS}.
\end{abstract}

\section{Introduction and preliminaries}
\label{sec:introduction}

Some ten years ago quantum (metric) graphs were proposed as a
convenient paradigm for the study of quantum chaos in compact
\cite {KS} and scattering \cite {KSscat} systems. The crucial
point which highlighted the close similarity between metric graphs
- whose only claim to complexity is their topology - and chaotic
Hamiltonian flows is the formal similarity between the trace
formulae \cite {Roth,Gutzwiller} which express the spectral
densities as sums over periodic orbits. Requiring additionally
that the lengths of the bonds are rationally independent, and that
the graph is well connected, render the spectrum of the
Schr\"odinger operator on graphs sufficiently disordered to
display spectral statistics which are consistent with the
predictions of Random Matrix Theory. Another important contact
point was the identification of the classical dynamics on the
graph, which is derived from the quantum evolution, and which
describes, under certain conditions, an exponential approach to
equilibrium, in analogy with mixing Hamiltonian flows. The $\zeta
$ function for the Peron-Frobenius operator on the graph, can be
written in terms of the same periodic orbits which are used for
the quantum spectral $\zeta$ function - in close similarity with
the expansion of the Ruelle $\zeta$ function as a sum over
periodic orbits for chaotic Hamiltonian flows. The derivation of
the trace formula for quantum graphs which was presented in \cite
{KS} differs from the original method \cite {Roth}. It uses
another approach (called sometimes the ``scattering approach"),
which reveals in a natural way the underlying classical dynamics.

Discrete graphs, where only the graph \emph{topology} and not its
\emph{metric} plays a r\^ ole are mostly studied in number theory,
combinatorics \emph{etc}. There is abundant literature relating to
various aspects of graphs. Much of the relevant material to the
present discussion can be found in \cite {Chung,AudreyBook}.
Audrey Terras' review \cite{AudreyRev} surveys the field, and its
relation to quantum chaos. The present work attempts to highlight
further this quantum chaos connection, by proposing  trace
formulae and spectral $\zeta$ functions, and linking them with the
Ihara $\zeta$ function \cite{Ihara} and some of its recent
generalizations. To introduce these concepts, a few preliminaries
and definitions are necessary, and they are provided below.

A graph $\mathcal{G}$ consists of $V$ vertices connected by $B$
bonds. The $V\times V$ \emph{connectivity} (or \emph{adjacency})
matrix $C$ is defined such that $C_{i,j}=1(0)$ if the vertices
$i,j$ are connected (disconnected). Graphs with parallel bonds or
loops are excluded. The \emph{valency} (some times referred to as
the \emph{degree}) of a vertex is the number of bonds which
emanate from a vertex. It is denoted by $v_i=\sum_{j=1}^V
C_{i,j}$. To any bond $b=(i,j)$ one can assign an arbitrary
direction, resulting in two \emph{directed bonds}, $d=(i,j)$ and
$\hat d =(j,i)$. Thus, the graph can be viewed as $V$ vertices
connected by bonds $b=1,\cdots,B$ or by $2B$ directed bonds
$d=1,\cdots,2B$. (The notation $b$ for bonds and $d$ for directed
bonds will be kept throughout). It is convenient to associate with
each directed bond $d =(j,i)$ its \emph{origin} $o(d) =i$ and
\emph{terminus} $t(d)=j$ so that $d$ points from the vertex $i$ to
the vertex $j$. The bond $d'$ follows $d$ if $t(d)=o(d')$. A
periodic orbit (cycle) of length $n$ is a sequence of $n$
successively following directed bonds $d_1,\cdots,d_n$  and $d_1$
follows $d_n$. Cyclic permutations of the bonds generate the same
periodic orbit. A primitive periodic orbit is an orbit which
cannot be written as a repetition of a shorter periodic orbit.
 The set of primitive $n$-periodic orbits will be denoted by
$\mathcal{P}(n)$, and $\mathcal P =
\bigcup_{n=2}^{\infty}\mathcal{P}(n)$. An important subset of
$\mathcal{P}(n)$ is the set of $n$ primitive periodic orbits
without back-scatter, namely, periodic orbits where $d_{i+1} \ne
\hat d_{i}$. The corresponding sets will be denoted by $\mathcal
{C}(n)$ and $\mathcal C = \bigcup_{n=2}^{\infty}\mathcal{C}(n)$.

 The Laplacian of a discrete graph is defined as
 \begin{equation}
 L\equiv-C+D,
 \label{eq:laplacian}
\end{equation}
where $C$ is the connectivity matrix, and $D$ is a diagonal matrix
with $D_{i,i}=v_i$. It is a self-adjoint operator whose spectrum
consists of $V$ non negative real numbers. The spectrum is
determined as the zeros of the secular function (characteristic
polynomial)
\begin{equation}
Z_L(\lambda) \equiv\det (\lambda I^{(V)}-L)\ .
 \label{eq:ZsubV}
\end{equation}
Here, $\lambda$ is the spectral parameter and $I^{(V)}$ is the
unit matrix in $V$ dimensions. The lowest eigenvalue is $0$, and
it is simple if and only if the graph is connected.

It is sometimes convenient to generalize the Laplacian
(\ref{eq:laplacian}) by replacing the matrix $C$ by a matrix
$\tilde C$ whose zero entries coincide with those of $C$, but
arbitrary, strictly positive weights $w_{i,j}\ (= w_{j,i})$
replace the values $1$ where $C_{i,j} =1$. One then defines
$\tilde D_{i,i}\equiv u_i=\sum_{j}\tilde C_{i,j}$ and the
generalized Laplacian is
\begin{equation}
 \tilde L\equiv-\tilde C+\tilde D.
 \label{eq:glaplacian}
\end{equation}
The spectrum of $\tilde L$ consists of the zeros of the secular
equation (characteristic polynomial) $Z_{\tilde{L}}(\lambda)\equiv
\det(\lambda I^{(V)}-\tilde{L})$. The spectrum is  non negative,
$0$ is in the spectrum and it is a simple eigenvalue if and only
if the graph is connected.

The focus of the present work is on $\zeta$ functions and trace
formulae for  discrete graphs. This research subject was initiated
by Ihara \cite {Ihara} who defined a $\zeta$ function for a graph
as
 \begin{equation}
\zeta(u)^{-1} \equiv \prod_{n} (1-u^n)^{|\mathcal{C}(n)|}\ ,
 \label{eq:iharadef}
 \end{equation}
where $|\mathcal{C}(n)|$ is the cardinality of the set
$\mathcal{C}(n)$, and $u\in \mathbb{C}$ with $|u|$  sufficiently
small to ensure the convergence of the infinite product. Following
Ihara's original work, several authors (see e.g., \cite {Hurt} for
a survey of the methods) have proved that
\begin{equation}
\zeta(u)^{-1} = (1-u^2)^{r-1}\det(I^{(V)}-u C +u^2 Q)\ .
 \label{eq:ihara}
 \end{equation}
Here, $r\equiv B-V+1$ is the \emph{rank} of the graph (the number
of independent cycles on the graph or equivalently, the rank of
its fundamental group). $I^{(V)}$ is the unit matrix in $V$
dimensions, $C$ is the connectivity matrix, and the diagonal
matrix  $Q\equiv D-I^{(V)}$. If the graph is $v$-regular, that is
$v_i =v \ \ \forall i$, the non trivial poles of the Ihara $\zeta$
(the trivial poles are at $u = \pm 1$) can be easily computed from
the eigenvalues of the graph Laplacian (\ref{eq:laplacian}).

The following $\zeta$ function defined by H.M. Stark \cite
{Starkmultipath} will serve as an example of the more recent
developments in the field. Consider a matrix $Y$ in the space of
directed bonds
\begin{equation}
Y_{d',d} \equiv  \eta_{d',d} \ \delta_{o(d'),t(d)}\
(1-\delta_{d',\hat d}).
\end{equation}
where  $\eta_{d',d}$ are arbitrary. Note that matrix elements
between reversed bonds are excluded. Associate with any primitive
periodic orbit $c\in \mathcal{C}$ the amplitude
\begin{equation}
f_c \equiv \eta_{d_n,d_{n-1}}\ \eta_{d_{n-1},d_{n-2}},\ \cdots\
\eta_{d_2,d_1}\ \eta_{d_1,d_n}\ .
\end{equation}
Then,
 \begin{equation} \zeta_E(Y)^{-1}\equiv \prod_{c\in
\mathcal{C}}\left(1-f_c\right ) = \det(I^{(2B)}-Y),
\label{eq:stark}
\end{equation}
where $I^{(2B)}$ is the unit matrix in $2B$ dimensions. This
result will be used in the last section.

In the next section, other $\zeta$ functions  are defined,
discussed, and expressed as rational functions which are
reminiscent of (\ref{eq:ihara}) and (\ref{eq:stark}), but  are
different in many respects. Trace formulae for the spectra of the
Laplacians (\ref {eq:laplacian},\ref{eq:glaplacian}) will also be
derived. In the last section, the approach developed here will be
compared with its analogues in the theory of quantum graphs, and
the ``classical dynamics" on the discrete graph will be proposed.

\section{Secular functions, $\zeta$ functions and trace formulae}
\label{sec:secularzeta}

To start, an alternative form of the secular equations for the
Laplacians (\ref {eq:laplacian},\ref{eq:glaplacian}) will be
derived.  It is convenient to begin with a detailed derivation for
the traditional Laplacian (\ref {eq:laplacian}). The necessary
modifications  for the generalized form will be indicated later.
For both Laplacians, the secular function will be shown to take
the form
\begin{equation}
Z_S(\lambda) = \frac{1}{2^B}\left (\det U(\lambda)\right
)^{-\frac{1}{2}}\det \left (I^{(2B)}-U(\lambda)\right )
 \label{eq:secular}
\end{equation}
where $U(\lambda)$ is a unitary matrix of dimension $2B$ which
depends on the spectral parameter $\lambda$. By construction,
$Z_S(\lambda)$ is real for $\lambda \in \mathbb{R}$, and its zeros
will be shown to coincide (with their multiplicity) with the
spectrum of the Laplacian. Thus $Z_S(\lambda)$ and $Z_L(\lambda)$
can differ at most by a multiplicative function of $\lambda$ which
does not vanish for real $\lambda$. This construction of the
secular function paraphrases  the ``scattering approach"
introduced in \cite {KS} for quantum graphs. (Derivations which
are similar in spirit were discussed in \cite{Novikov, Cataneo},
see also \cite {Kuchment} and references cited therein).

To compute an eigenvector   $\psi=(\psi_1,\cdots,\psi_V)$ of $L$,
corresponding to an eigenvalue $\lambda$, the following steps are
taken. To each bond $b =(i,j)$ one associates a \emph{bond wave
function}
\begin{equation}
\psi_b(x) = a_b\ {\rm e}^{i\frac{\pi}{4}x} + a_{\hat b}\ {\rm
e}^{-i\frac{\pi}{4}x} \ \ \ ,
 \ \ x\in \{\pm1\}
 \label{eq:wfunction}
 \end{equation}
subject to the condition
\begin{equation}
\psi_b(1)= \psi_i \ \ \ , \ \ \ \psi_b(-1)= \psi_j  \ .
\label{eq:bcond}
 \end{equation}
Consider any vertex indexed by $i$, and the  bonds $(b_1,b_2,
...b_{v_i})$ which emanate  from $i$. The corresponding bond wave
functions have to satisfy three requirements in order to form a
proper eigenvector of $L$.

\noindent {\it I. Uniqueness}: The value of the eigenvector at the
vertex $i$, $\psi_i$, computed in terms of the bond wave functions
is the same for all the bonds emanating from $i$. The following
$v_i-1$ independent equalities express this requirement.
 \begin{equation}
 \hspace{-15mm}
 a_{b_1}\ {\rm e}^{i\frac{\pi}{4}} + a_{\hat b_1}\ {\rm
e}^{-i\frac{\pi}{4}} = a_{b_2}\ {\rm e}^{i\frac{\pi}{4}} + a_{\hat
b_2}\ {\rm e}^{-i\frac{\pi}{4}} =\ \cdots \  = a_{b_{v_i}}\ {\rm
e}^{i\frac{\pi}{4}} + a_{\hat b_{v_i}}\ {\rm e}^{-i\frac{\pi}{4}}\
.
 \label {eq:uniq}
\end{equation}

\noindent {\it II. $\psi$ is an eigenvector of $L$}  : At the
vertex $i$, $\sum_{j=1}^{v_i} L_{i,j}\psi_j=\lambda \psi_i$. In
terms of the bond wave functions this reads,
\begin{equation}
\hspace{-10mm}
 -\sum_{l=1}^{v_i}\left [a_{b_l}\ {\rm
e}^{-i\frac{\pi}{4} } + a_{\hat b_l}\ {\rm e}^{+i\frac{\pi}{4}
}\right ]
 =(\lambda - v_i)\ \frac{1}{v_i} \sum_{m=1}^{v_i}
\left [ a_{b_m}\ {\rm e}^{i\frac{\pi}{4}} + a_{\hat b_m}\ {\rm
e}^{-i\frac{\pi}{4}}\right ]\ .
 \label{eq:lapleq}
 \end{equation}
To get the equation above, $\psi_i$ was presented as
\begin{equation}
\psi_i = \frac{1}{v_i}\sum_{j=1}^{v_i}\left(a_{b_j}\ {\rm
e}^{i\frac{\pi}{4}} + a_{\hat b_j}\ {\rm
e}^{-i\frac{\pi}{4}}\right) \ .
\end{equation}
 Together, (\ref{eq:uniq}) and (\ref{eq:lapleq}) provide
$v_i$ homogeneous linear relations  between  the $2v_i$
coefficients $a_d$, where $d$ stand for directed bonds which are
either incoming to ($t(d)=i$) or outgoing from ($o(d)=i$) the
vertex $i$. Using these equations, the outgoing coefficients are
expressed in terms of the incoming ones,
 \begin{equation}
 \label{eq:scatmat}
a_d =\sum_{d'\ :\ t(d')=i}  \sigma^{(i)}_{d,d'}(\lambda) \ a_{d'}\
\ \
   \ \forall \   d \ :\ o(d)=i\ ,
\end{equation}
where,
\begin{eqnarray}
 \label{eq:scatmatdetail}
 \hspace {-5mm}
\sigma^{(i)}_{d,d'}(\lambda )&=& i\left(\delta_{\hat d,
d'}-\frac{2}{v_i}\frac{1}{1-i(1-\frac{\lambda}{v_i})} \right) \ =
i\left(\delta_{\hat d, d'}-\frac{1}{v_i} (1+{\rm e}^{i\alpha_i
(\lambda)} )\right)\nonumber \\
\hspace {-5mm} {\rm e}^{i\alpha_i(\lambda)}\ \ \ &=&
\frac{1+i(1-\frac{\lambda}{v_i})} {1-i(1-\frac{\lambda}{v_i})}\ .
\end{eqnarray}
The \emph{vertex scattering matrices} $\sigma^{(i)}(\lambda)$ are
the main building blocks of the present approach. They distinguish
clearly between back-scatter transitions ($\hat d = d'$) and the
transitions to other bonds, for which the same strength is given,
independently of the original and the final bonds. For real
$\lambda$ the vertex scattering matrices are unitary matrices and
they are the discrete analogues of the vertex scattering matrices
derived for the Schr\"odinger equation on graphs \cite {KS}.

\noindent {\it III. Consistency} : The linear relation between the
incoming and the outgoing coefficients (\ref {eq:scatmat}) must be
satisfied simultaneously at all the vertices. However, a directed
bond $(i,j)$ when observed from the vertex $j$ is \emph{outgoing},
while when observed from $i$ it is \emph{incoming}. This
consistency requirement is implemented  by introducing the
\emph{Evolution Operator} $U_{d'.d}(\lambda)$ in the $2B$
dimensional space of  directed bonds,
\begin{equation}
 U_{d',d}(\lambda) = \delta_{t(d),o(d')}\
 \sigma^{(t(d))}_{d',d}(\lambda)\ .
 \label{eq:umatrix}
\end{equation}
($U$ is also referred to in the literature as the \emph{Bond
Scattering Matrix} \cite{KS}). The evolution operator is unitary $
U\ U^{\dagger} = I^{(2B)}$ for $\lambda\in \mathbb{R}$ due to the
unitarity of its constituents $\sigma ^{(i)}$. Denoting by $\bf a$
the $2B$ dimensional vector of the directed bonds coefficients
$a_d$ defined above, the consistency requirement reduces to,
\begin{equation}
U(\lambda)\ {\bf a} = {\bf a}\ .
\end{equation}
This can be satisfied only for those values of $\lambda$ for which
\begin{equation}
\xi(\lambda)\ \equiv \ \det\left (I^{(2B)}-U(\lambda)\right )\ = \
0 \ .
\end{equation}
For real $\lambda$ the spectrum of $U(\lambda)$ is restricted to
the unit circle. Therefore $|\xi(\lambda)|$ is finite for all
$\lambda \in \mathbb{R}$. Due to (\ref {eq:scatmatdetail}) the
matrix elements of $U(\lambda)$ are ratios of monomials in
$\lambda$. These two properties imply that
$\xi(\lambda)=p(\lambda)/q(\lambda) $ where $p$ and $q$ are
polynomials of the same degree in $\lambda$, and their degree is
at most $2B$. The zeros of $q(\lambda)$ coincide with the poles of
$\det U(\lambda)$. They are complex because $|\det U(\lambda)|=1$
for $\lambda\in \mathbb{R}$. A straight forward computation
yields,
\begin{equation}
\hspace{-15mm}
 \det U(\lambda) = \prod_{j=1}^V\
 \frac{1+i(1-\frac{\lambda}{v_j})}{1-i(1-\frac{\lambda}{v_j})} \ ,
 \ \ \Rightarrow  \ \ q(\lambda) = Const \ \prod_{j=1}^V\ \left(
 1-i(1-\frac{\lambda}{v_j})\right ) \ .
  \label{eq:detU}
\end{equation}
Thus, $\det U$ has exactly $V$ complex poles, implying that the
degree  of $p(\lambda)$ which equals the degree of  $q(\lambda)$
is also $V$. Note finally that the zeros of $p(\lambda)$ coincide
with the zeros of the secular function $Z_L(\lambda) =\det
(\lambda I^{(V)}-L)$ which is also a polynomial of degree $V$.
Hence, $p(\lambda)$ and $Z_L(\lambda)$ are identical up to a
constant factor. It is convenient to define the secular equation
so that it is real on the real axis. This can be achieved by
multiplying $\xi(\lambda)$ by $(\det U(\lambda))^{-\frac{1}{2}}$.
A further factor of $2^{-B}$ normalizes the resulting function to
approach $1$ as $|\lambda|\rightarrow \infty$. The resulting
secular equation reads
\begin{eqnarray}
\hspace{-20mm}
 Z_S(\lambda)\ &=& \ \frac{1}{2^B}\left (\det
U(\lambda)\right )^{-\frac{1}{2}}\det \left
(I^{(2B)}-U(\lambda)\right ) \\
\hspace{-20mm} &=& \frac{1}{2^B} \prod_{j=1}^V\
\left(\frac{1+i(1-\frac{\lambda}{v_j})}
{1-i(1-\frac{\lambda}{v_j})}\right)^{\frac{1}{2}}
\frac{p(\lambda)}{q(\lambda)}\  =
 \ \frac{\det\ (\lambda I^{(V)} -L)}
{\prod_{j=1}^V (v_j^2+(v_j-\lambda)^2)^{\frac{1}{2}}} \ .\nonumber
 \label{eq:zsubS}
\end{eqnarray}
This expression for the secular equation is the basis for the
further results of the present work. To begin, use is made of the
fact that the spectrum of $U(\lambda)$ for $\mathcal{I}m(\lambda)
< 0$ is confined to the interior of the unit circle. Thus, for for
any $\lambda$ with an arbitrarily small (but finite) negative
imaginary part, we expand
\begin{equation}
\log \det(I^{(2B)}-U(\lambda))= -\sum_{n=1}^{\infty}\frac{1}{n}\
{\rm tr}
 U^n(\lambda)\ ,
 \label{eq:logdet}
\end{equation}
and
 \begin{equation}
{\rm tr} U^n(\lambda) = \sum_{m : m|n}\ m\sum_{p\in
\mathcal{P}(m)}a_p(\lambda)\ .
 \label{eq:truton}
 \end{equation}
The sum above is over all the primitive periodic orbits $p$ with
period $m$ which is a divisor of $n$, $p=d_1,\ \cdots\ ,d_m$ and
 \begin{equation}
 a_p(\lambda) =\sigma_{d_{1},d_m} (\lambda)\cdots\sigma_{d_{2},d_1} (\lambda) \ .
 \label{eq:amplitude}
\end{equation}
The explicit dependence of  $a_p(\lambda)$ on $\lambda$ is
obtained from the following expressions for the vertex scattering
matrix elements,
\begin {eqnarray}
\label{eq:explisig}
 \hspace{-10mm} \sigma_{d',d} = \left \{
\begin{array}{lr}
\ \ \ \left [ \frac{4}{v_j^2+(v_j-\lambda)^2}\right
]^{\frac{1}{2}}\ \ \ {\rm e}^{\ i
 [\arctan(1-\frac{\lambda}{v_j})]/2} & {\rm for} \
\ \ d' \ \ne \ \hat d \ , \\
\left
[1-\frac{4(v_j-1)}{[v_j^2+(v_j-\lambda)^2]}\right]^{\frac{1}{2}}
{\rm
e}^{-i\arctan\frac{2(v_j-\lambda)}{(v_j-1)^2+(v_j-\lambda)^2-1}}\
& {\rm for} \ \ \ d' \ = \ \hat d \ ,
\end{array}
 \right .
\end{eqnarray}
where $  j = t(d) =  o(d')$. The explicit expressions above were
written so that for real $\lambda$ the absolute square of the
$a_p$ is a product of ``transition probabilities", while the phase
of $a_p$ is a sum which plays the r\^ole of the ``action" or
``length" associated with the periodic orbit.  Substituting (\ref
{eq:truton}) in (\ref {eq:logdet}), and summing over the
repetition numbers $\frac {n}{m}$ one gets,
\begin{equation}
\det(I^{(2B)}-U(\lambda)) = \prod_{p\in
\mathcal{P}}(1-a_p(\lambda))\ . \label{eq:pprod}
\end{equation}
The $\zeta$ function which is introduced in the present work is
defined as
\begin{equation}
\zeta_S(\lambda)^{-1}\ = \ \prod_{p\in{\cal P}}(1-a_p(\lambda))\ .
\label{eq:defzeta}
\end{equation}
Combining  (\ref{eq:zsubS}) and (\ref{eq:pprod}) with the
definition of $\zeta_S(\lambda)$ gives
\begin{equation}
\zeta_S(\lambda)^{-1} \ =\
 \frac{\det(\lambda I^{(V)}- L)}
{\prod_{j=1}^V\left (v_j+i(v_j- \lambda )\right )}\ .
 \label{eq:zetavsz}
\end{equation}
This is one of the main results of the present work. It provides a
``Ihara" - like identity which expresses an infinite product over
primitive periodic orbits on the graph in terms of the
characteristic polynomial of the graph discrete Laplacian. The
main difference is that here, all the periodic orbits, including
orbits with back-scattering,  contribute to the product, and that
the amplitudes $a_p$ depend on the spectral parameter in a more
complicated way. To get a closer look at the $\zeta_S$ function
and its relation to the Ihara $\zeta$ function, it is instructive
to write $\zeta_S$ for a general $v$-regular graph. For this
purpose, it is convenient to define a new complex variable,
\begin{equation}
z=\frac{1+i(1-\frac{\lambda}{v})}{1-i(1-\frac{\lambda}{v})}
\label{eq:zdef}
\end{equation}
which is a $1 \leftrightarrow 1$ map  of $\mathbb{R}$ to the unit
circle in $\mathbb{C}$. With these simplifications, $\zeta_S(z)$
(\ref {eq:zetavsz}) reduces to
\begin{equation}
\zeta_S(z)^{-1}= \left(\frac{2z}{z+1}\right)^V\ \det\left (C +
iv\frac{z-1}{z+1}\ I^{(V)}\right ) \ .
 \label{eq:simplezeta}
\end{equation}
It is convenient to define $\gamma_S(z)=z^{\frac{V}{2}}
\zeta_S(z)$, in terms of which a functional equation for $\zeta_S$
can be written as
\begin{equation}
\gamma_S(z^{-1}) = \left(\gamma_S(z^\ast)\right)^\ast\ ,
 \label{eq:functional}
\end{equation}
where $(\cdot)^\ast$ stands for complex conjugation. Functional
equations of similar type are satisfied also by the Ihara $\zeta$
function (for $v$-regular graphs) as well as by most other
functions of this genre. Typically, functional equations enable
the analytical continuation of $\zeta$ functions which are defined
by infinite products, beyond their radius of convergence. Here
also it provides the analytic continuation of $\zeta_S(z)$ to the
exterior of the unit disc.

The periodic amplitudes $a_p(z)$ simplify considerably for
$v$-regular graphs. Denote by $n_p$ the period of the primitive
periodic orbit $p$, and by $\beta_p$ the number of vertices in $p$
where back-scattering occurs: $\beta_p = \sharp \left \{i\ :\ d_i=
\hat d_{i+1} ,\  d_i \in p,\ d_{n_p+1}=d_1\right \}$. Then,
\begin{equation}
a_p(z)= {\rm e}^{-i\frac{\pi}{2}n_p}\left(\frac
{1+z}{v}\right)^{n_p-\beta_p}\ (-1)^{\beta_p}\ \left( 1-
\frac{1+z}{v} \right)^{\beta_p}\ .
\end{equation}
\vspace {10mm}

The results above pave the way to the derivation of trace formulae
for the discrete Laplacians. Trace formulae provide a powerful
tool in spectral theory.  They express the spectral density
(written down formally  as a sum of Dirac $\delta$ functions
located at the spectral set) in terms of information derived from
the manifold metric. The spectral density is written as a sum of
two contributions - both of which have a geometric origin. The
first is a smooth function of $\lambda$ whose asymptotic limit at
$\lambda \rightarrow \infty$ was first studied by Weyl. The second
contribution is an infinite sum over periodic geodesics on the
manifold. The equality between the spectral density and its
geometric representation should be understood only in the sense of
distributions. An analogous trace formula will be derived now for
the discrete Laplacian. Making use of Cauchy theorem and the fact
that $Z_S(\lambda)$ is analytic in the vicinity of the real
$\lambda$ axis, and real on it, one can write,
\begin{eqnarray}
\label{eq:cauchy}
d(\lambda) &=&\sum_{j=1}^V \delta(\lambda-\lambda_j)\\
&=& \frac{1}{\pi}\ \lim_{\epsilon\rightarrow 0^+}{\mathcal Im}
\frac{{\rm d}\ }{{\rm d}\lambda} \log Z_S(\lambda - i\epsilon) \ .
\end{eqnarray}
Using
\begin{equation}
Z_S(\lambda)\ = \ \frac{1}{2^B}\left (\det U(\lambda)\right
)^{-\frac{1}{2}}\det \left (I^{(2B)}-U(\lambda)\right ) \ ,
\end{equation}
the explicit form of $\det U(\lambda)$ (\ref{eq:detU}) and the
periodic orbit expansion (\ref {eq:truton}), one gets,
\begin{equation}
\hspace{-15mm}
 d(\lambda)= \frac{1}{\pi} \sum_{j=1}^V
\frac{1}{v_j}\ \frac{1}{1+(1-\frac{\lambda}{v_j})^2} \ - \
\frac{1}{\pi} {\mathcal Im}\ \frac {{\rm d}\ }{{\rm d}\lambda}
\sum_{r=1}^{\infty}\ \sum_{p\in {\mathcal P}}\ \frac{1}{n(p)}
|a_p(\lambda)|^r {\rm e}^{i r \phi_p(\lambda)} \ .
 \label {eq:traceformula}
\end{equation}
The first term is the ``smooth" (Weyl) contribution to the
spectral density. It consists of a sum of Lorenzians with poles at
$\lambda_j=v_j(1 \pm i)$. This sum is analogous to Wigner's
semi-circle density in Random Matrix Theory.

The explicit expression for the fluctuating part can be written
down explicitly using (\ref {eq:explisig}). Noting  that the
$a_p(\lambda)$ are complex numbers with $\lambda$ dependent phases
$\phi_p(\lambda)$, the periodic orbit sum in the trace formula is
a fluctuating function of $\lambda$. It is the term which turns
the r.h.s. of (\ref {eq:traceformula}) to a distribution when
$\epsilon \rightarrow 0$.

So far, the discussion was restricted to the ``traditional"
Laplacians. The extension to the generalized Laplacians, starts by
modifying the definition of the bond wave functions (\ref
{eq:wfunction}) to read,
\begin{equation}
\psi_b = {\sqrt w_b}(a_b{\rm e}^{i\frac{\pi}{4}x} +a_{\hat b}{\rm
e}^{-i\frac{\pi}{4}x})\ .
\end{equation}
Then, following the same steps as above, the vertex scattering
matrices are derived, and they take the form
\begin{equation}
\label{eq:gscatmatdetail} \hspace{-10mm} \tilde
\sigma^{(i)}_{d,d'} (\lambda)= i\left(\delta_{\hat d,
d'}-\frac{1}{u_i} (1+{\rm e}^{i\alpha_i (\lambda)}\ )\sqrt{w_d
w_{d'}} \right) \ \ ; \ \ {\rm e}^{i\alpha_i(\lambda)}=
\frac{1+i(1-\frac{\lambda}{u_i})} {1-i(1-\frac{\lambda}{u_i})}\ ,
\end{equation}
where $u_j=\sum_j w_{i,j}$ as defined previously. The subsequent
derivation follows the same steps, resulting in the generalized
$\zeta_S$ function,
\begin{equation}
\zeta_{\tilde S}(\lambda)^{-1}\ \equiv \ \prod_{p\in{\cal
P}}(1-a_p(\lambda))\ =\
 \frac{\det(\lambda I^{(V)}- \tilde L)}
{\prod_{j=1}^V\left (u_j+i(u_j- \lambda )\right )}\ .
\end{equation}
A trace formula is also derived in the same way,
\begin{eqnarray}
\hspace{-15mm} d(\lambda) &=&\sum_{j=1}^V \delta(\lambda-\tilde
\lambda_j)
 =  \frac{1}{\pi}\ \lim_{\epsilon\rightarrow 0^+}{\mathcal I}m
\frac{{\rm d}\ }{{\rm d}\lambda} \log Z_{\tilde S}(\lambda -
i\epsilon) \\
&=& \frac{1}{\pi} \sum_{j=1}^V \frac{1}{u_j}\
\frac{1}{1+(1-\frac{\lambda}{u_j})^2}\ \ - \ \ \frac{1}{\pi}
{\mathcal Im}\ \frac {{\rm d}\ }{{\rm d}\lambda}
\sum_{r=1}^{\infty}\ \sum_{p\in {\mathcal P}}\  \frac{1}{n(p)}
 |a_p| ^r\ {\rm e}^{i r  \phi_p(\lambda)}\nonumber
\end{eqnarray}
The expressions for $a_p(\lambda)$ can be derived by a simple
modification of (\ref {eq:explisig})  and therefore they will not
be written down here.

\section{Classical dynamics}
\label{sec:summary}

The present approach emerges from the alternative secular function
for the spectrum of Laplacians, based on the  quantum evolution
operator $U(\lambda)$ in the space of directed bond amplitudes
${\bf a}\in l^2(\mathbf{C}^{2B}) $. Consider $U(\lambda)$ as a
quantum map which maps this $2B$ dimensional space onto itself.
$U$ is unitary and hence the map conserves the $l^2$ norm - the
quantum probability. The condition $U(\lambda){\bf a}={\bf a}$ can
be interpreted as a requirement that $\lambda_n$ is an eigenvalue
if there exists a non trivial vector ${\bf a}$ which is stationary
under the action of the quantum map \cite{gnuzus}. The requirement
of stationarity is naturally associated with the eigenvalue being
in the spectrum of the underlying Hamiltonian.

The building blocks for the theory are the vertex scattering
matrices. Similar matrices appear in the theory of quantum graphs.
There, they emerge when the Schr\"odinger equation on the graph is
augmented by vertex boundary conditions which render the resulting
operator self adjoint. The self adjoint extension is not unique,
and depends the spectral parameter $k$ and on an arbitrary
parameter $\kappa$ which interpolates between the ``Dirichlet"
($\kappa=0$) and the ``Neumann" ($\kappa=\infty$) boundary
conditions \cite {KS,Schraeder}. The scattering matrices for
discrete graphs are obtained from their quantum graph analogues by
replacing $\kappa/k$ by $\lambda$.

The unitary quantum evolution operator is the starting point for
the construction of a classical evolution on the discrete graph.
The classical ``phase space" in this case are the probability
vectors ${\bf \rho}\in l^2(\mathbf{R}^{2B})$ where the components
of ${\bf \rho}$ are interpreted as the probabilities to find the
classical system on the corresponding directed bonds. The
classical transition matrix is constructed from the quantum
probability to make a transition from $d$ to $d'$
\begin{equation}
 M_{d',d}=|U_{d',d}|^2\ .
\end{equation}
The unitarity of $U$ implies that $M$ is bi-stochastic, namely,
$\sum_d M_{d',d}=\sum_{d'} M_{d',d}=1$. This transition matrix
induces a discrete, random walk dynamics in phase space. If $n$
denotes the discrete ``time",
\begin{equation}
{\bf \rho}(n+1) = M {\bf \rho}(n)\ .
\end{equation}
This Markovian evolution preserves the $l^1$ norm - the classical
probability. The spectrum of $M$ is confined to the interior of
the unit circle. $1$ is always an eigenvalue corresponding to an
eigenvector with equal components which describes the system in an
equilibrated state. When the eigenvalue $1$ is the only eigenvalue
on the unit circle, the classical dynamics drives the system to
equilibrium at a rate which depends on the distance of the next
highest eigenvalue to the unit circle. This classical dynamics is
identical to the one which was introduced in the study of quantum
graphs \cite{KS}. It plays an important r\^ole in the theory of
spectral statistics on quantum graphs
\cite{KS,Tanner,Berkolaiko,SGAA}. Finally, the analogue of the
Ruelle $\zeta$ function for the evolution induced by $P$ can be
easily written down starting with the secular function
\begin{equation}
Z_M(\mu) \equiv \det (I^{(2B)}-\mu M) \label{eq:clasec}
 \end{equation}
The periodic orbit sum is identical to the trace formula (\ref
{eq:traceformula}) in which the amplitudes $a_p$ are replaced by
their absolute squares.

To emphasize the intricate connections between the concepts
developed here and their predecessors \cite{Ihara,Stark, Sunada},
consider a $v$-regular graph ($v>2$), and the classical evolution
operator obtain for the spectral parameter $\lambda= v+i(v-2)$,
corresponding to $z=v-1$ in (\ref{eq:zdef}). At this value,
$\sigma_{d,\hat d}=0$ and  $\sigma_{d',d} =1$ for $d'\ne \hat d$.
The resulting classical evolution matrix $M^{\sharp}$ needs to be
multiplied by $(v-1)^{-1}$ to make it a legitimate (probability
conserving) evolution operator. The resulting evolution does not
permit back-scatter, and therefore, the secular equation
(\ref{eq:clasec}) can be computed using Stark's $\zeta$ function
(\ref {eq:stark}), with $Y =\frac{\mu}{v-1}M^{\sharp}$. The
product over the set of non back-scattering primitive periodic
orbits  becomes identical to the one appearing in the Ihara zeta
function (\ref {eq:iharadef}). Using (\ref{eq:ihara}), one finally
gets,
\begin{eqnarray}
\label {eq:connect}
 Z_{M^{\sharp}}(\mu)&= & \prod_{n}\left(1-
(\frac{\mu}{v-1})^n\right)^{|\mathcal{C}(n)|} \\
  & = &\left(1-
(\frac{\mu}{v-1})^2\right)^{r-1}\det \left(I^{(V)}(1+
\frac{\mu^2}{v-1})-\frac{\mu}{v-1}C\right)\nonumber \ .
\end{eqnarray}
Thus, the spectrum $m_j$ of  $M^{\sharp}$ consists of $r-1$ fold
degenerate eigenvalues at $m_j^{(\pm)} = \pm \frac {1}{v-1}$, and
the rest which can be computed from the spectrum of the discrete
Laplacian $\lambda_j$
\begin{equation}
m_j^{\pm} = \frac{(v-\lambda_j)\pm
\sqrt{(v-\lambda_j)^2-4(v-1)}}{2(v-1)} \ .
\end{equation}
The eigenvalue $0$ of the Laplacian corresponds to the eigenvalues
$1$ and $\frac{1}{v-1}$ of $M^{\sharp}$. The gap in the classical
evolution spectrum is determined by the first non zero eigenvalue
of $L$.

The comment above may have interesting and novel consequences
going beyond its anecdotal appearance. In quantum graphs, one can
choose vertex scattering matrices from a much larger variety than
offered by the vertex scattering matrices (\ref
{eq:scatmatdetail}). Thus, it is possible to construct vertex
scattering matrices which do not scatter backwards, but with equal
scattering probability to the other vertices. The unitarity is
maintained by a proper choice of the phases of the scattering
amplitudes \cite{Harrison}. In such cases, and for $v$-regular
graphs, the classical analogues are identical with $M^{\sharp}$
and (\ref{eq:connect}) is applicable. Working with such systems is
particularly interesting because in quantum chaos, the gap between
the eigenvalue 1 and the rest of the spectrum determines whether
the spectrum of the $U$ matrix (and hence of the Schr\"odinger
operator) display the statistics predicted by Random Matrix
Theory, in the limit of large graphs. For non back-scattering
dynamics (\ref {eq:connect}) reduces the problem to the study of
the spectrum of the Laplacian. The behavior of the gap in the
laplacian spectrum of large graphs is an important subject in the
theory of discrete graphs and number theory, related amongst
others to the Ramanujan conjecture \cite{ramanujan}. A detailed
discussion of this connection will take the present manuscript far
afield, and it is deferred to a future publication.

 \vspace
{10mm}

\noindent {\bf Acknowledgments}

 \noindent It is a pleasure to thank A Terras, M Kotani,  H Stark
and T Sunada  for introducing me to the fascinating world of the
discrete graphs, and for many discussions and suggestions. The
comments and suggestions offered by P Kuchment, S Gnutzmann, I
Oren and R Band are highly appreciated. J. Harrison's help in
computing a few examples of vertex scattering matrices without
back-scatter is also acknowledged. This work was supported by the
Minerva Center for non-linear Physics, the Einstein (Minerva)
Center at the Weizmann Institute and EPSRC grant 531174. I am
indebted to the Isaac Newton Institute and Clare Hall for their
hospitality while much of this work was carried out.


\vspace {2cm}


 {\bf Bibliography}
\begin{thebibliography}{10}

\bibitem{KS} Tsampikos Kottos and Uzy Smilansky,
\emph{Quantum Chaos on Graphs}, Phys. Rev. Lett.  {\bf 79},4794-
4797, (1997). and  \emph{Periodic orbit theory and spectral
statistics for quantum  graphs}, Annals of Physics {\bf 274},
76-124 (1999).

\bibitem{Ihara} Y. Ihara, \emph{On discrete subgroups of the two by two
projective linear group over a p-adic field}, J. Mat. Soc. Japan,
{\bf 18} (1966), 219-235.


\bibitem{Stark} H. M. Stark and A. A. Terras, \emph{Zeta Functions of
Finite Graphs and Coverings}, Adv. in Math. {\bf 121}, (1996)
124-165.

\bibitem{Sunada} Motoko Kotani and Toshikazu Sunada,\emph{ Zeta
Functions on Finite Graphs}, J. Math. Sci. Univ. Tokyo {\bf 7}
 no. 1, 7-25, (2000).

\bibitem{KSscat} Tsampikos Kottos and Uzy Smilansky,
 \emph{ Chaotic Scattering on Graphs} , Phys. Rev. Lett., {\bf 85}
968, (2000).


\bibitem{Roth} J. P. Roth in ``Lecture notes in Mathematics:
Theorie de Potential" (A. Dold and B. Eckmann, Eds.) p. 521,
Springer Verlag, New-York/Berlin (1985).

\bibitem{Gutzwiller} M.C. Gutzwiller, J. Math. Phys. {\bf 12} 343
(1971).

\bibitem{Chung}  Fan R. K. Chung, \emph{Spectral Graph Theory}, Regional
Conference Series in Mathematics  {\bf 92},  American mathematical
Society(1997).

\bibitem{AudreyBook} A. A. Terras, \emph{Fourier Analysis on Finite
Groups and Applications} London Mathematical Society Student Texts
{\bf 43} Cambridge University Press, Cambridge UK (1999).

\bibitem{AudreyRev} A. A. Terras, \emph{Arithmetic Quantum Chaos},
IAS/Park City Mathematical Series {\bf 12} 2002 333-375


\bibitem {Hurt}Hurt, N.E. \emph{The prime geodesic theorem and quantum
mechanics on finite volume graphs:a review.} Rev. Math. Phys. {\bf
13}, (2001), 1459-1503.

\bibitem{Starkmultipath} H.M. Stark, \emph{Multipath zeta functions of
graphs}. preprint (2007)

\bibitem{Novikov} S. Novikov, \emph{Schr\"odinger operators on graphs
and symplectic geometry. }Field Inst. Communications {\bf 24}
(1999), 397-413.

\bibitem{Cataneo} Carla Cattaneo, \emph{The Spectrum of the Continuous
Laplacian on a Graph.} Mh. Math. {\bf 124} 215-235 (1997).

\bibitem{Kuchment} Peter Kuchment, \emph{Quantum Graphs: I. Some basic
structures.} Waves in random Media {\bf 14} (2004), S107-S128.

\bibitem{gnuzus}Sven Gnutzmann and Uzy Smilansky,
\emph{Quantum Graphs: Applications to Quantum Chaos and Universal
Spectral Statistics.} Advances in Physics {bf 55} (2006) 527-625.

\bibitem {Schraeder} V. Kostrykin and R Schrader, \emph{Kirchoff's Rule
for quantum wires}, J. Phys. A. Math. Gen. {\bf 32}(1999) 595-630.

\bibitem {Tanner} G. Tanner, J. Phys. A {\bf 34},8484 (2001)
and {\it ibid} A{\bf 35},5985 (2002).

\bibitem {Berkolaiko} G. Berkolaiko, \emph{Spectral gap of doubly
stochastic matrices generated from equidistributed unitary
matrices},  J. Phys. A {\bf 34},(2001) 319-326.

\bibitem {SGAA} S.Gnutzmann and A. Altland, Phys. Rev. Lett. {\bf
93} 194101 (2004).

\bibitem{ramanujan} A. Lubotzky, R. Phillips and P. Sarnak,
\emph{Ramanujan graphs}, Combintorica {\bf 8} (1988), 261-277.

\bibitem{Harrison} J. Harrison, private communication.

\end{thebibliography}
\end{document}